\documentclass[a4paper,twocolumn,11pt]{quantumarticle}
\pdfoutput=1
\usepackage[utf8]{inputenc}
\usepackage[english]{babel}
\usepackage[T1]{fontenc}
\usepackage{amsmath}

\usepackage{lipsum}

\usepackage{graphicx}
\usepackage{dcolumn}
\usepackage{bm}
\usepackage{hyperref}
\usepackage{cleveref}
\usepackage{xcolor}
\usepackage{physics}
\usepackage{multirow}
\usepackage{orcidlink}
\usepackage{amsfonts}
\usepackage{subcaption}
\usepackage{dsfont}
\usepackage[normalem]{ulem}
\usepackage{pgfplots}
\usepackage{amssymb}
\usepackage{cite}

\usepackage{tikz}

\usepackage[numbers,sort&compress]{natbib}
\PassOptionsToPackage{compress}{natbib}

\let\emph\relax
\DeclareTextFontCommand{\emph}{\itshape}

\def\Ltar{L_{\mathrm{tar}}}

\newcommand{\beq}{\begin{align}}
\newcommand{\eeq}{\end{align}}

\begin{document}
\title{Mapping twist fields to local operators via tensor networks}

\author{Andrea Bulgarelli\orcidlink{0009-0002-2917-6125}}\email{abulgare@uni-bonn.de}
\affiliation{Transdisciplinary Research Area ``Building Blocks of Matter and Fundamental Interactions'' (TRA Matter) and Helmholtz Institute for Radiation and Nuclear Physics (HISKP), University of Bonn, Nussallee 14-16, 53115 Bonn, Germany}

\author{Marco Panero\orcidlink{0000-0001-9477-3749}}\email{marco.panero@unito.it}
\affiliation{Department of Physics, University of Turin, Via Pietro Giuria 1, I-10125 Turin, Italy}
\affiliation{INFN, Turin division, Via Pietro Giuria 1, I-10125 Turin, Italy}

\author{Paolo Stornati\orcidlink{0000-0003-4708-9340}}\email{pstornat@bsc.es}
\affiliation{Barcelona Supercomputing Center, Pla\c{c}a Eusebi G\"uell, 1-3, 08034 Barcelona, Spain}

\author{Luca Tagliacozzo\orcidlink{0000-0002-5858-1587}}\email{luca.tagliacozzo@csic.es}
\affiliation{Instituto de Fisica Fundamental IFF-CSIC, Calle Serrano 113b, 28006 Madrid, Spain}
\affiliation{Quantum Advanced Research Center (QuARC), CSIC, Calle Serrano 113b, 28006 Madrid, Spain}

\begin{abstract}
Twist fields are a powerful formal tool to compute R\'enyi entropies in quantum many-body systems, but their conventional formulation in tensor network states involves operations acting on virtual degrees of freedom, which are not directly accessible in experiments. In this work, we construct explicit local operators acting on the physical Hilbert space whose expectation values reproduce the action of twist fields in matrix product states. Our construction is exact in the injectivity limit and when the tensor is chosen at the center of orthogonality, and provides a direct operational method to evaluate R\'enyi entropies without accessing auxiliary tensor indices. We test our formulation numerically in the transverse-field Ising model, demonstrating rapid convergence to the exact entanglement entropy as the injectivity scale is reached. Furthermore, we show that twist operators determined from relatively small reference systems can be reliably transferred to larger systems, once the reference size exceeds a characteristic scale set by the correlation length. Since the resulting operators admit a decomposition in terms of a finite number of local observables, our results provide a scalable and experimentally accessible framework to probe entanglement in quantum simulators.
\end{abstract}

\maketitle
\section{Introduction}
\label{sec:introduction}

Quantifying entanglement in quantum many-body systems has become one of the central challenges of modern physics since it provides a unified framework to understand colletive phenomena, from the geometry of space time, to the characterization of the phases of quantum matter~\cite{Takayanagi:2025ula, Zeng:2015pxf}. R\'enyi entropies are a family of entanglement monotones that interpolate between the von Neumann entropy and the purity, encode universal information about quantum critical phenomena, and are directly sensitive to the topological and symmetry structure of quantum phases~\cite{Renyi:1961omo, Vidal:1998re, Calabrese:2004eu, Levin:2006zz, Kitaev:2005dm}. Given a subsystem $\mathcal{A}$ and its complement $\mathcal{B}$, R\'enyi entropies are defined as
\begin{align}
    S_n(\mathcal{A}) = \frac{1}{1-n}\log\Tr\rho_{\mathcal{A}}^n, \qquad
    \rho_{\mathcal{A}} = \Tr_{\mathcal{B}}\rho.
\end{align}
Over the years, they have progressively evolved from a purely theoretical tool into a target of experimental detection: recent experiments with ultracold atoms, trapped ions, and Rydberg arrays demonstrated that it is possible to directly access low-order R\'enyi entropies via randomized measurements, many-body interference, and shadow tomography protocols~\cite{Islam:2015mom, Brydges:2019wut, Huang:2020tih, Elben:2020hpu}. These advances 
are stimulating increasing interest in designing 
experimental protocols to measure such entropies, 
which 
are both theoretically exact and experimentally implementable at scale.

From a theoretical perspective, for
a quantum many-body system or a quantum field theory defined in one spatial dimension 
the natural setting 
to study R\'enyi entropies 
associated with the degrees of freedom of a subsystem $\mathcal{A}$ (that, for definiteness, we take to be a a finite spatial interval) 
is based on
the expression of 
the path integral of the
theory on a replicated Riemann surface, with 
different sheets connected through a cut 
along $\mathcal{A}$. This has been proven to be equivalent to the insertion of quasi-local fields, called ``twist fields''~\cite{Dixon:1986qv, Calabrese:2004eu, Cardy:2007mb, Swingle:2010jz, Hung:2014npa, Doyon:2025xvo} at the end of the interval; in particular, 
in a 
conformal field theory (CFT), the $n$-th R\'enyi entropy 
is proportional to the two-point function of branch-point twist fields with scaling dimension $\Delta_n = (c/12)(n - 1/n)$, where $c$ is the central charge~\cite{Calabrese:2004eu}.

Matrix-product states (MPS) and higher-dimensional tensor networks (TN), such as projected entangled-pair states (PEPS), provide a natural numerical framework to study these quantities in lattice systems~\cite{Fannes:1990ur, Vidal:2003pmm, Perez-Garcia:2006nqo, Verstraete:2008cex}. Through the density-matrix renormalization group (DMRG)~\cite{White:1992zz, White:1993zza}, MPS give essentially exact ground states for gapped one-dimensional Hamiltonians, and remain efficient even near criticality when the gap closes only polynomially with the system size \cite{Verstraete:2006mdr, Tagliacozzo:2007rda}. The entanglement entropy of a contiguous subsystem is directly accessible from the Schmidt decomposition of the MPS at negligible additional cost. However, R\'enyi entropies of disjoint or non-contiguous subsystems are not encoded in the standard MPS Schmidt structure, and their computation requires a different approach.

\begin{figure*}[t!]
    \centering
    \begin{subfigure}{.35\linewidth}
    \includegraphics[width=1\linewidth]{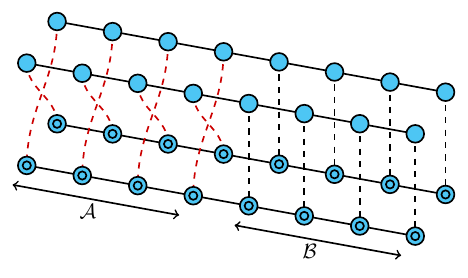}
    \caption{}
    \end{subfigure}
    \hspace{1.2cm}
    \begin{subfigure}{.35\linewidth}
    \includegraphics[width=1\linewidth]{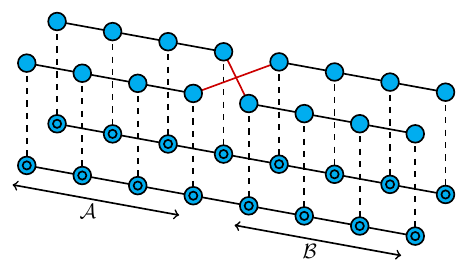}
    \caption{}
    \label{subfig:virtual_contraction}
    \end{subfigure}
    \caption{(a) Swap operator on the physical bonds of the MPS. (b) Equivalent contraction with a swap at the virtual level. Dashed lines represent physical indices, and tensors with a circle inside are the complex conjugate.}
    \label{fig:virtual_real_contraction}
\end{figure*}

The twist field formalism was extended to MPS in ref.~\cite{Coser:2013qda}, where it was shown that
$k$-interval  $n$-R\'enyi entropies can be expressed as 
$2k$-point correlation functions of twist fields implemented as generalized transfer matrices in a replicated MPS. These transfer matrices are built from
$n$ copies of the standard MPS transfer matrix, concatenated with a cyclic swap operator acting on the \emph{virtual} bond indices. Such a virtual swap operator acts as a local operator in the tensor network, exactly as the twist field does in field theories.

This construction is elegant and computationally powerful, and can be used to efficiently extract the CFT data~\cite{Coser:2013qda}, but it has a fundamental limitation from an experimental standpoint: the swap operator acts on the auxiliary degrees of freedom of the tensor network, which have no direct physical interpretation and are inaccessible to any measurement performed on the quantum state itself. Thus, in experiments one has to use physical swap operators which extend for the full length of the interval whose entanglement has to be characterized. 

In the present work, we solve this problem by exploiting the notion of \emph{injectivity} in 
MPS~\cite{Perez-Garcia:2010maz, Molnar:2018hls}.
An MPS is injective at a given blocking scale $l$ when the map from the virtual Hilbert space to the physical Hilbert space defined by a block of $l$ consecutive tensors is injective — that is, if operations on the virtual level can be faithfully represented by physical operators. Injectivity is a generic property of MPS: in particular, for any state with finite correlation length $\xi$, it is reached exponentially fast 
in $l/\xi$~\cite{Perez-Garcia:2010maz}. 
In the injectivity limit, the virtual swap operator that defines the MPS twist field can be explicitly lifted to the physical level by inverting the blocked MPS tensor via singular-value decomposition (SVD). The resulting operator acts on $2l$ physical sites across two replicas and reproduces the R\'enyi entropy exactly.

Crucially, the physical twist operator we construct admits a decomposition as a linear combination of a finite number of local Pauli strings. The number of independent coefficients is controlled by the symmetries of the construction and grows only moderately with the blocking length $l$. Since Pauli string expectation values can be estimated from local measurements and few-body correlators, this means that R\'enyi entropies of arbitrary multi-interval subsystems become accessible from a \emph{restricted and scalable} set of physical measurements, without requiring full state tomography or reconstruction of the reduced density matrix.

We test the construction numerically in the transverse-field Ising model, across a broad range of couplings, demonstrating convergence to exact R\'enyi entropies that is rapid away from criticality and systematic near the critical point, where larger blocking lengths are required to reach the injectivity scale. We further demonstrate that twist operators determined from small reference systems can be reliably transferred to significantly larger systems, once the reference size exceeds a characteristic scale proportional to the correlation length. This transferability makes the protocol scalable and directly suited to current quantum simulation platforms, including Rydberg atom arrays, trapped ions, and superconducting qubit systems.

This article is organized as follows. In section~\ref{sec:tensor_network_states_and_Renyi_entropies}, we introduce MPS, R\'enyi entropies, and the transfer matrix formalism. Section~\ref{sec:injectivity_and_twist_fields} discusses the construction of physical twist operators in the injectivity limit and at the orthogonality center. Numerical benchmarks are presented in section~\ref{sec:numerical_tests}, while section~\ref{sec:approximate_schemes} analyzes the transferability of twist operators across system sizes. Section~\ref{sec:experimental_accessibility_of_the_twist_operator} addresses experimental accessibility and the Pauli decomposition, and section~\ref{subsec:higher_dimensions_PEPS} extends the analysis to PEPS and two-dimensional systems. Finally, our conclusions are summarized in section~\ref{sec:conclusions}. Throughout this article, unless otherwise stated, we express the dimensionful quantities in units of the appropriate power of the lattice spacing.

\section{Tensor network states and R\'enyi entropies}
\label{sec:tensor_network_states_and_Renyi_entropies}

Tensor network states are quantum many-body states described by the contraction of a network of elementary tensors. They provide efficient variational \emph{Ans\"atze} for ground states of strongly correlated systems. The most prominent example is given by MPS, which lie at the core of the 
DMRG algorithm~\cite{White:1992zz, White:1993zza}. 
The latter has become the standard numerical method for solving one-dimensional gapped Hamiltonians, and remains efficient even for systems approaching criticality, provided the closing of the gap with the system size $L$ is only polynomially fast.

Given the central role they play, here we focus on MPS. These describe the state of a one-dimensional system of size $L$ as the contraction of $L$ rank-three tensors, as sketched here:
\begin{equation}
\nonumber
\begin{tikzpicture}[baseline=(current bounding box.center),scale=0.8]
\definecolor{mpsblue}{rgb}{0.0, 0.65, 0.9}
    \draw[thick, black] (1.1,0.8) -- (7.7,0.8);

    \foreach \x in {1,...,7}{
        \draw[thick, black] (1.1*\x,0.8) -- (1.1*\x,-0.18);
    }

    \foreach \x in {1,...,7}{
        \draw[thick, fill=mpsblue] (1.1*\x,0.8) circle (0.30);
    }

    \node[scale=0.9] at (3.9,1.25) {$\chi$};
    \node[scale=1.0] at (6.30,-0.21) {$d$};

\end{tikzpicture}
\end{equation}
Each tensor has two auxiliary (or virtual) legs, of dimension $\chi$, which are contracted to build the physical wavefunction, and a physical leg of dimension $d$, 
which carries the information about 
the local Hilbert space. MPS have a long history in the study of entanglement in quantum many-body systems. Through a simple ``gauge'' transformation, one can directly extract the Schmidt eigenvalues associated with any bipartition into adjacent constituents, at essentially the same leading computational cost ($\chi^3$) required to obtain the ground state itself via DMRG. This has made MPS a natural framework for investigating entanglement scaling at and off criticality.


However, in many cases one is interested in the entanglement between disjoint regions, or equivalently in the R\'enyi entropies of subsystems composed of non-consecutive sites. This setting has been explored for instance in ref.~\cite{Coser:2013qda}, where it was shown that, analogously to the situation in quantum field theory, $k$-interval $n$-R\'enyi entropies can be obtained within the MPS formalism from $2k$-point correlation functions of twist fields. These twist fields are implemented by inserting generalized transfer matrices (TM) in the replicated system, defined as tensor products of $n$ copies of the standard MPS transfer matrix concatenated with a swap operator acting on the auxiliary spaces, see fig.~\ref{fig:virtual_real_contraction}. This naturally raises the question, whether such generalized TM can be realized as TM encoding the action of  local operators defined on the replicated physical system, rather than on the auxiliary degrees of freedom, which are not accessible in experiments. 

When moving to higher dimensions, the natural generalization of MPS are tensor product states (TPS), also called projected entangled pair states (PEPS). Indeed a PEPS  quantum many-body wavefunction is represented by a network of local tensors placed on the sites of the lattice and connected through auxiliary, or virtual, indices. The physical amplitudes are obtained by contracting all virtual indices, while the dimension of these auxiliary spaces controls the amount of entanglement that can be represented. This construction was introduced in the quantum-information language by Verstraete and Cirac as a variational \emph{Ansatz} for quantum many-body systems in two and higher dimensions~\cite{Verstraete:2004cf}. Closely related tensor-product variational states had already been developed in the context of classical statistical mechanics and two-dimensional quantum systems by Nishino and collaborators~\cite{Nishino:2000zoq,Nishino:2000ygc,nishio2004}. In this sense, PEPS combine the tensor-product variational ideas of these earlier works with the entanglement-based perspective that made matrix product states so successful in one dimension.

\subsection{Bond dimension and correlation length}

In an MPS with finite bond dimension $\chi$, correlation functions decay exponentially with distance. For translation-invariant systems in the thermodynamic limit, the rate of decay is set by the correlation length $\xi$, determined by the two leading eigenvalues $t_1,t_2$ of the MPS transfer matrix  $E = \sum_i A_i A_i^{\dagger}$. Assuming $t_1\ge t_2$,  the correlation length can be expressed as $\xi = 1/\log(t_1/t_2)$. In special cases exact degeneracies of the TM spectrum can lead to a lack of decay of specific correlation functions, but here we restrict to lattice systems whose low-energy physics is governed by a Lorentz-invariant critical point, i.e. a conformal field theory. In this case Calabrese and Cardy established that the entanglement entropy of a half-chain scales as $S \propto \tfrac{c}{6}\log \xi$, with $c$ the central charge of the CFT~\cite{Calabrese:2004eu}. As a consequence, the bond dimension of an MPS approximating such a state is bounded from below by $\chi \gtrsim \xi^{c/6}$. 
Complementary results~\cite{Verstraete:2006mdr} show that $\chi$ is also bounded from above by the exponential of the scaling of R\'enyi entropies of order $n<1$. For our purposes, it is sufficient 
to state that, when the correlation length is increased, a corresponding increase in bond dimension is required, in order 
to faithfully capture the ground state.

\section{Injectivity and twist fields}
\label{sec:injectivity_and_twist_fields}

Closely related to the bond dimension and the correlation length is the notion of \emph{injective} MPS~\cite{Perez-Garcia:2006nqo}, where the local tensors define injective maps from the auxiliary Hilbert spaces to the physical degrees of freedom. Injectivity is typically reached 
by coarse-graining blocks of sites of order the correlation length~\cite{Verstraete:2004qk}. The significance of injectivity is that operations on the auxiliary level can be faithfully represented by appropriate operators on the physical Hilbert space, rendering the description locally invertible. The larger the correlation length, the larger the block one must coarse-grain before reaching an injective description. 
In our discussion, 
we will often assume local injectivity of the MPS. Even though this approximation becomes 
less and 
less accurate as $\xi$ grows, one can always redefine the tensor network by blocking more tensors. For fixed correlation length, the injectivity limit is reached exponentially fast in the size of the blocks.

\subsection{Inversion of the MPS}

With injectivity established, we can return to the earlier question regarding whether a local physical operator can implement the twist field. For injective MPS, the answer is affirmative: the explicit construction of twist fields in ref.~\cite{Coser:2013qda} at the level of virtual bonds can be lifted to the physical Hilbert space by coarse-graining until injectivity is achieved. At that point, the equivalence between virtual and physical operators ensures that one can 
explicitly identify a 
local operator on the replicated system that acts as the twist field. This observation underpins the construction that we will 
discuss 
in the following.

Without loss of generality, we 
focus on 
the construction of the twist field for $n=2$ replicas. 
The MPS tensors are assumed to have bond dimension $\chi$ and physical dimension $d^l$, where the integer $l\geq 1$ is tuned to achieve injectivity. The central idea 
in our approach consists in inverting 
the MPS tensor, seen as a $\chi^2 \times d^l$ matrix, from the physical index, in order to access the virtual legs. Given a tensor $A$, its inverse $\bar{A}$ is constructed using singular value decomposition, as shown here:\footnote{For a similar construction in a different context see also refs.~\cite{ljubotina2024,petrova2025}.}

\begin{equation}
\nonumber
\def\x{4.5}
\def\y{1.5}
\def\ys{8}
\def\xx{13}
\def\yy{3}
\def\xf{3}
\def\yf{1.5}
\begin{tikzpicture}[baseline=(current bounding box.center),scale=0.7]
\filldraw[fill=cyan!70!blue!10!, draw=black] (0,0) circle (1cm);
\draw[ultra thick, color=black] (1,0) -- (2,0);
\draw[ultra thick, color=black] (-1,0) -- (-2,0);
\draw[ultra thick, color=black] (0,-1) -- (0,-2);
\draw (2,.5) node { $\chi$};
\draw (-2,.5) node { $\chi$};
\draw (.4,-1.8) node { $d^l$};
\draw (0,0) node {\Large $A$};

\draw (3,-.5) node {\Large $=$};

\filldraw[fill=green!20!, draw=black] (\x,\y) -- (\x+2,\y) -- (\x+1,\y-1) -- cycle;
\draw (\x+1,\y-.4) node {\Large $U$};
\draw[ultra thick, color=black] (\x+.5,\y-.5) -- (\x-.5,\y-.5);
\draw[ultra thick, color=black] (\x+.5+2,\y-.5) -- (\x-.5+2,\y-.5);

\draw[ultra thick, color=black] (\x+1,\y-1) -- (\x+1,\y-1.65);
\fill[color=black] (\x+1,\y-1.65) circle (.15cm);
\draw (\x+1.7,\y-1.5) node {\Large $S$};
\draw[ultra thick, color=black] (\x+1,\y-1.8) -- (\x+1,\y-1.8-.5);

\filldraw[fill=orange!20!, draw=black] (\x+1,\y-1.8-.5) -- (\x+2,\y-1.8-.5-1) -- (\x,\y-1.8-.5-1) -- cycle;
\draw (\x+1,\y-1.8-.5-.6) node {\Large $V^\dagger$};
\draw[ultra thick, color=black] (\x+1,\y-1.8-.5-1) -- (\x+1,\y-1.8-.5-2);

\draw (\x-.6,\y-.1) node { $\chi$};
\draw (\x+2+.6,\y-.1) node { $\chi$};
\draw (\x+1.5,\y-1.8-.3-2) node { $d^l$};

\end{tikzpicture}
\end{equation}

\begin{equation}
\nonumber
\def\x{4.5}
\def\y{1.5}
\def\ys{8}
\def\xx{13}
\def\yy{3}
\def\xf{3}
\def\yf{1.5}
\begin{tikzpicture}[baseline=(current bounding box.center),scale=0.7]
\filldraw[fill=cyan!70!blue!10!, draw=black] (0,-\ys) circle (1cm);
\draw[ultra thick, color=black] (1,-\ys) -- (2,-\ys);
\draw[ultra thick, color=black] (-1,-\ys) -- (-2,-\ys);
\draw[ultra thick, color=black] (0,-\ys+1) -- (0,-\ys+2);
\draw (2,-.5-\ys) node { $\chi$};
\draw (-2,-.5-\ys) node { $\chi$};
\draw (.4,1.8-\ys) node { $d^l$};
\draw (0,-\ys) node {\Large $\bar{A}$};

\draw (3,.5-\ys) node {\Large $=$};

\filldraw[fill=orange!20!, draw=black] (\x,\y-\ys+.5) -- (\x+2,\y-\ys+.5) -- (\x+1,\y-1-\ys+.5) -- cycle;
\draw (\x+1,\y-.4-\ys+.5) node {\Large $V$};
\draw[ultra thick, color=black] (\x+.5,\y-.5-1.8-\ys) -- (\x-.5,\y-.5-1.8-\ys);
\draw[ultra thick, color=black] (\x+.5+2,\y-.5-1.8-\ys) -- (\x-.5+2,\y-.5-1.8-\ys);

\draw[ultra thick, color=black] (\x+1,\y-1-\ys+.5) -- (\x+1,\y-1.65-\ys+.5);
\fill[color=black] (\x+1,\y-1.65-\ys+.5) circle (.15cm);
\draw (\x+1.9,\y-1.5-\ys+.5) node {\Large $S^{-1}$};
\draw[ultra thick, color=black] (\x+1,\y-1.8-\ys+.5) -- (\x+1,\y-1.8-.5-\ys+.5);

\filldraw[fill=green!20!, draw=black] (\x+1,\y-1.8-.5-\ys+.5) -- (\x+2,\y-1.8-.5-1-\ys+.5) -- (\x,\y-1.8-.5-1-\ys+.5) -- cycle;
\draw (\x+1,\y-1.8-.5-.6-\ys+.5) node {\Large $U^\dagger$};
\draw[ultra thick, color=black] (\x+1,\y+.5-\ys) -- (\x+1,\y+.5+1-\ys);

\draw (\x-.6,\y-1-1.8-\ys) node { $\chi$};
\draw (\x+2+.6,\y-1-1.8-\ys) node { $\chi$};
\draw (\x+1.5,\y-.6+2-\ys) node { $d^l$};

\end{tikzpicture}
\end{equation}

The contraction of the tensors $A$ and $\bar{A}$ from the physical leg is equivalent to the tensor product of two identity matrices, each one of dimension $\chi\times\chi$. This allows us to define a unitary operator whose expectation value reproduces the contraction in fig.~\ref{subfig:virtual_contraction}, namely

\begin{equation}\nonumber
\def\x{4.5}
\def\y{1.5}
\def\ys{8}
\def\xx{13}
\def\yy{3}
\def\xf{3}
\def\yf{1.5}
\def\yh{1}
\centering
\begin{tikzpicture}[baseline=(current bounding box.center),scale=0.6]


\filldraw[fill=blue!50!red!30!, draw=black] (\xx+1,-\ys+1-\yh) -- (\xx+1,-\ys-1-\yh) -- (\xx-1,-\ys-1-\yh) -- (\xx-1,-\ys+1-\yh) -- cycle;
\draw (\xx,-\ys-\yh) node {\Large $T$};
\draw[ultra thick, color=black] (\xx-.7,-\ys+1-\yh) -- (\xx-.7,-\ys+2-\yh);
\draw[ultra thick, color=black] (\xx+.7,-\ys+1-\yh) -- (\xx+.7,-\ys+2-\yh);
\draw[ultra thick, color=black] (\xx-.7,-\ys-1-\yh) -- (\xx-.7,-\ys-2-\yh);
\draw[ultra thick, color=black] (\xx+.7,-\ys-1-\yh) -- (\xx+.7,-\ys-2-\yh);

\draw (\xx+2.5,-\ys-\yh) node {\huge $=$};


\filldraw[fill=cyan!70!blue!10!, draw=black] (\xx+6,-\ys) circle (1cm);
\draw[thick, color=black] (-1+\xx+6,-\ys) to[out=180, in=180] (-1+\xx+6,-\ys-3);

\draw (\xx+6,-\ys) node {\Large $\bar{A}$};

\filldraw[fill=cyan!70!blue!10!, draw=black] (\xx+6,-\ys-3) circle (1cm);

\draw (\xx+6,-\ys-3) node {\Large $A$};


\filldraw[fill=cyan!70!blue!10!, draw=black] (\xx+6+\xf,-\ys+\yf) circle (1cm);
\draw[thick, color=black] (-1+\xx+6+\xf,-\ys+\yf) to[out=180, in=180] (-1+\xx+6+\xf,-\ys-3+\yf);

\draw (\xx+6+\xf,-\ys+\yf) node {\Large $\bar{A}$};

\filldraw[fill=cyan!70!blue!10!, draw=black] (\xx+6+\xf,-\ys-3+\yf) circle (1cm);

\draw (\xx+6+\xf,-\ys-3+\yf) node {\Large $A$};
\draw[ultra thick, color=red] (1+\xx+6,-\ys) to[out=0, in=0] (1+\xx+6+\xf,-\ys-3+\yf);
\draw[ultra thick, color=red] (1+\xx+6,-\ys-3) to[out=0, in=0] (1+\xx+6+\xf,-\ys+\yf);

\draw[ultra thick] (\xx+6,-\ys+1) -- (\xx+6,-\ys+2);
\draw[ultra thick] (\xx+6,-\ys-4) -- (\xx+6,-\ys-5);
\draw[ultra thick] (\xx+6+\xf,-\ys+1+\yf) -- (\xx+6+\xf,-\ys+2+\yf);
\draw[ultra thick] (\xx+6+\xf,-\ys-4+\yf) -- (\xx+6+\xf,-\ys-5+\yf);

\end{tikzpicture}
\end{equation}

The construction is not unique: a swap between sites $x$ and $x+1$ can be equivalently reproduced by an operator acting at $x$, as shown in the previous cartoon, or by one acting at $x+1$. 
Explicitly, if we denote the elements of $A$ as $A_{\mu\nu}^i(x)$, where Greek letters are associated with virtual indices, Latin letters with the physical ones, and the number in brackets labels the position of the tensor in the MPS, the ``forward'' twist operator is defined as the contraction
\begin{align}
    T^{(f)}_{ijkn}(x) = \bar{A}_{\mu\nu}^i(x) A_{\mu\sigma}^j(x) \bar{A}_{\rho\sigma}^k(x)  A_{\rho\nu}^n(x).
    \label{eq:twist_operator}
\end{align}

The ``backward'' twist operator $T^{(b)}(x+1)$ is obtained in a similar way from $A(x+1)$, by swapping the left-hand-side virtual legs. In the following, we only specify the superscript $(f)$ or $(b)$ when needed.

Finally, note that in the injectivity limit, the equality
\begin{align}
    \langle T \rangle = \Tr\rho_{\mathcal{A}}^2
    \label{eq:vev_twist_equal_Renyi}
\end{align}
is exact. We will explicitly show the validity of eq.~\eqref{eq:vev_twist_equal_Renyi} in the next section.

\section{Numerical tests}
\label{sec:numerical_tests}

\subsection{Results in 1D}
\label{subsec:results_in_1D}

In the present section, we numerically test eq.~\eqref{eq:vev_twist_equal_Renyi} in the specific case of the transverse field Ising model, with the Hamiltonian defined as
\begin{align}
    H = - \sum_{x=1}^{L-1}\sigma^z_x\sigma^z_{x+1} - g\sum_{x=1}^{L}\sigma^x_x,
\end{align}
where $\sigma^x$ and $\sigma^z$ 
denote 
Pauli matrices and open boundary conditions are assumed.

\begin{figure*}[t!]
    \centering
    \begin{subfigure}{.32\linewidth}
    \includegraphics[width=1\linewidth]{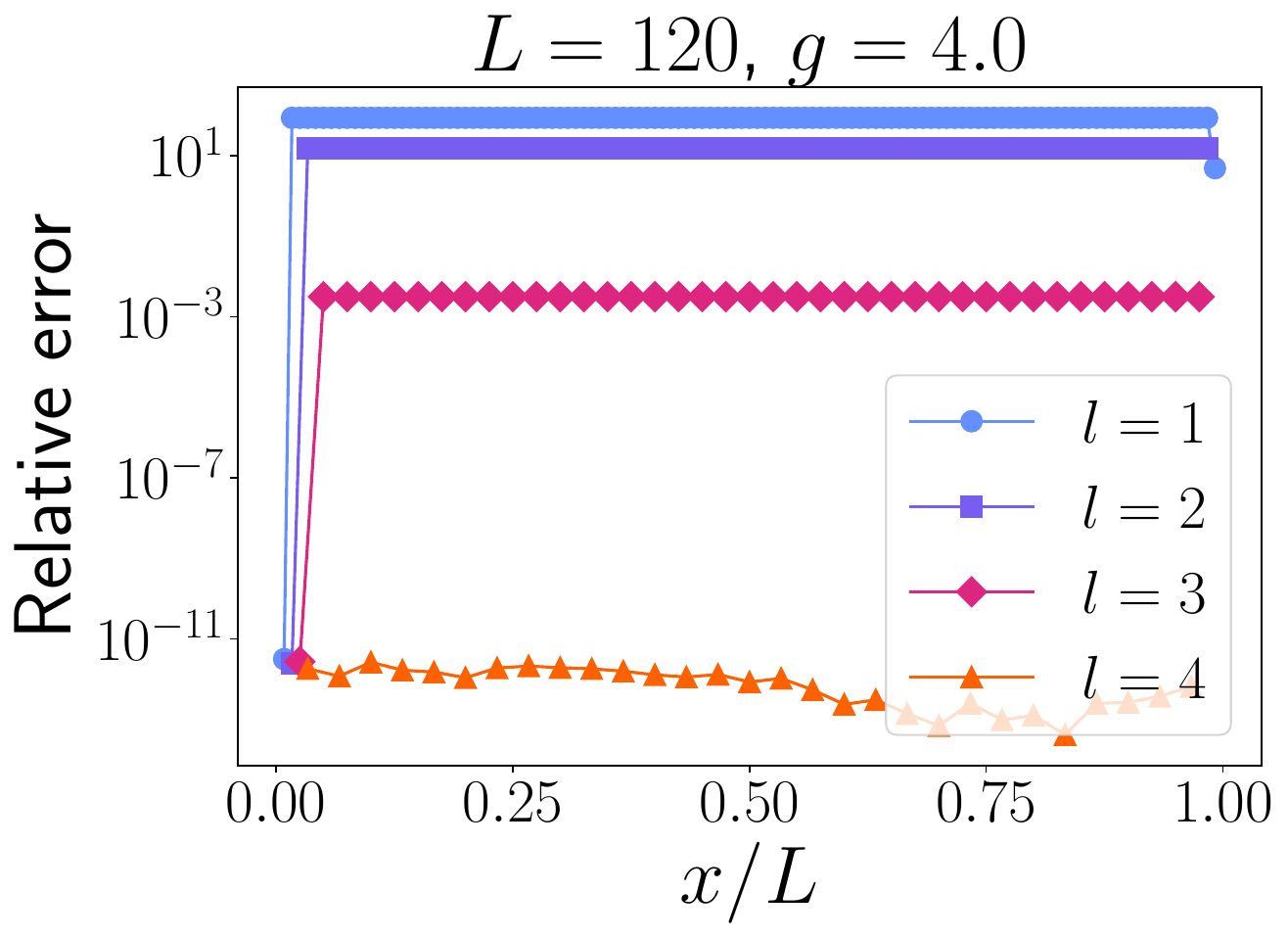}
    \caption{}
    \end{subfigure}
    \begin{subfigure}{.32\linewidth}
    \includegraphics[width=1\linewidth]{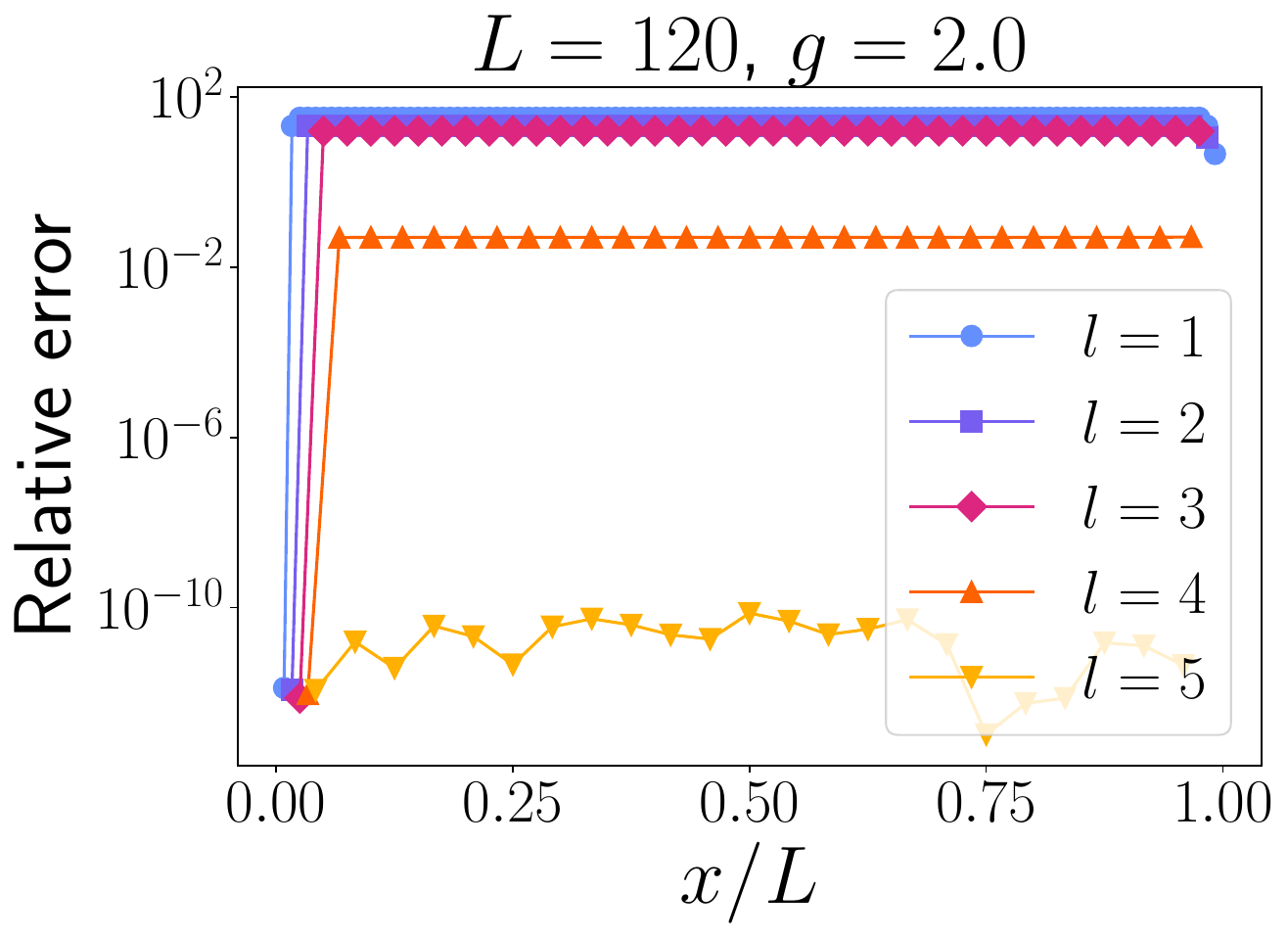}
    \caption{}
    \end{subfigure}
    \begin{subfigure}{.32\linewidth}
    \includegraphics[width=1\linewidth]{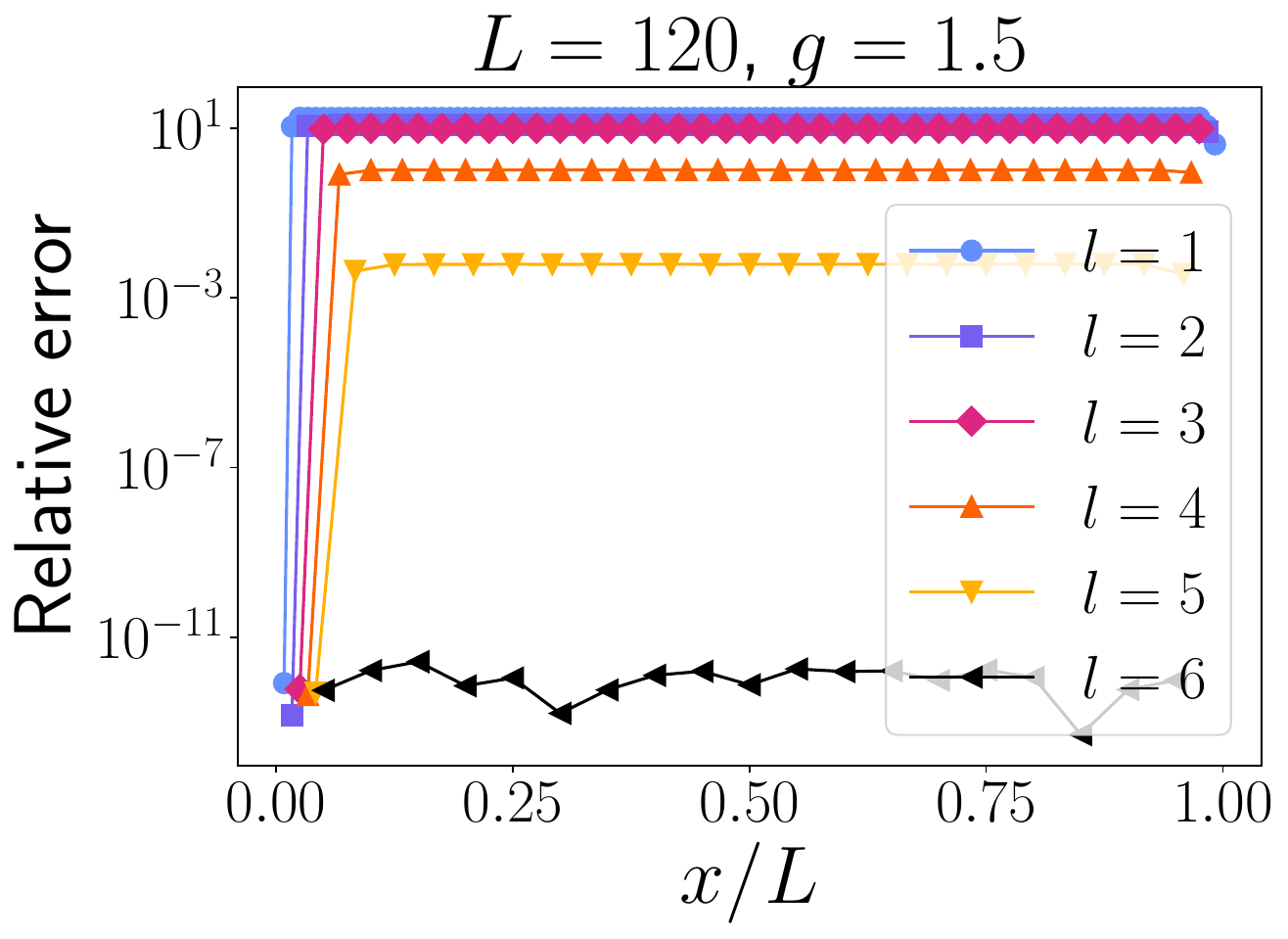}
    \caption{}
    \end{subfigure}
    \caption{Relative error 
    on the numerical estimate of 
    the second R\'enyi entropy 
    from the expectation value of the twist operator as the injectivity limit is approached. From left to right, the correlation length of the model increases, therefore the injectivity limit is obtained  
    with blocks of increasing sizes, 
    $l=4$, $5$, and $6$, for $g=4.0$, $2.0$, and $1.5$, respectively. 
    The results shown in these plots were obtained using 
    the forward definition of the twist operator.}
    \label{fig:approaching_injectivity_limit}
\end{figure*}

The results 
of our tests 
are shown in fig.~\ref{fig:approaching_injectivity_limit}, where we plot the relative error between the R\'enyi entropy and its approximation in terms of the expectation value of the twist operator, computed for 
increasingly 
larger blocks. Eventually, the result becomes exact. As the coupling $g$ approaches 
its critical value for this 
model, $g_c =1$, the injectivity limit is reached for increasingly larger block sizes.

The generalization to 
arbitrary values of 
$n$ is simply obtained contracting $n$ copies of $A$ with $n$ copies of $\bar{A}$, with the right-hand-side virtual legs (or left-hand-side ones in the backward construction) cyclically contracted. As anticipated, we constructed a local operator, acting on $l$ sites of each copy of the MPS, which computes the $n$-th R\'enyi entropy of a quantum spin chain.

\begin{figure*}[t!]
    \centering
    \begin{subfigure}{.4\linewidth}
    \includegraphics[width=1\linewidth]{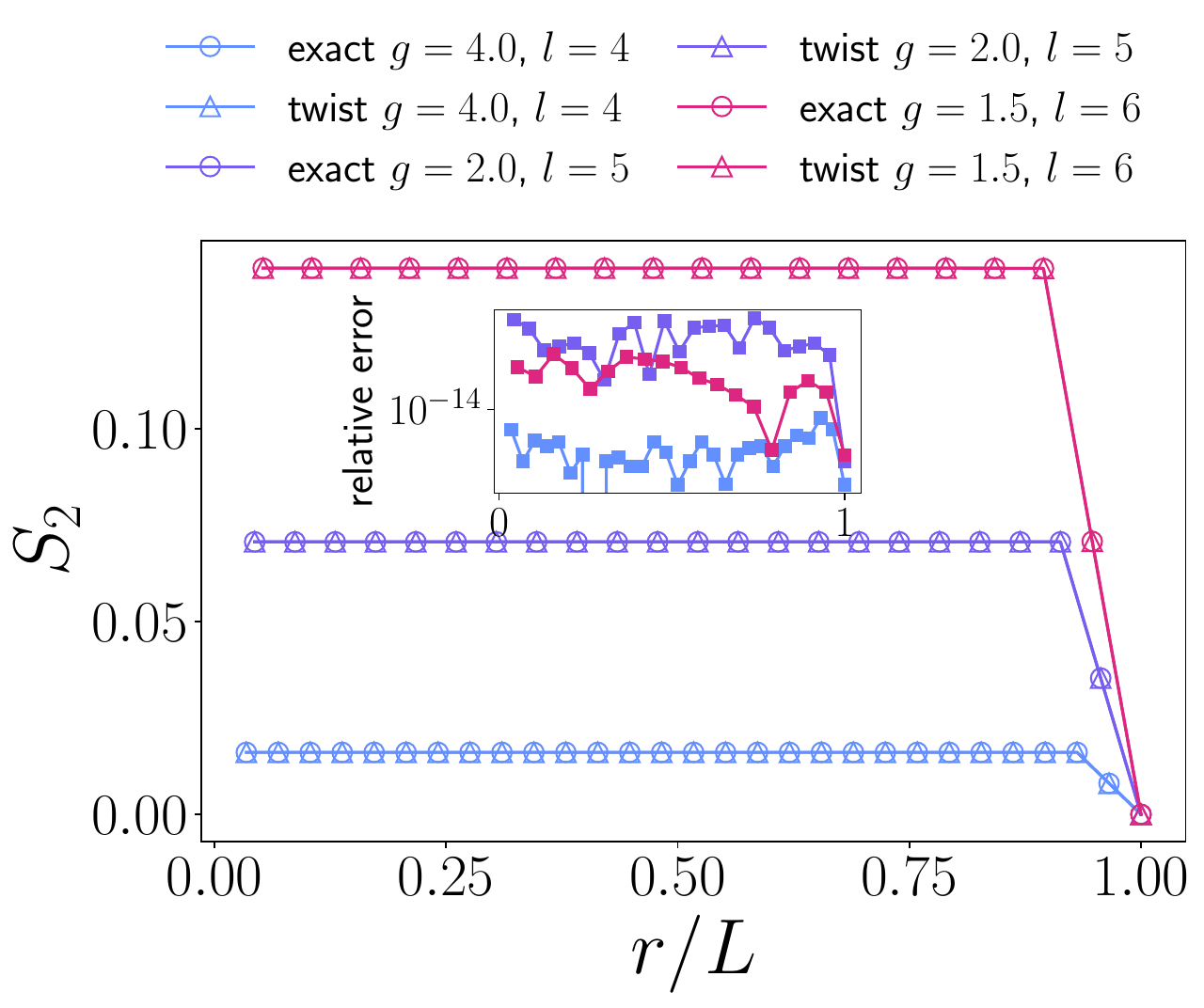}
    \caption{}
    \label{subfig:two_point_function}
    \end{subfigure}
    \begin{subfigure}{.4\linewidth}
    \includegraphics[width=1\linewidth]{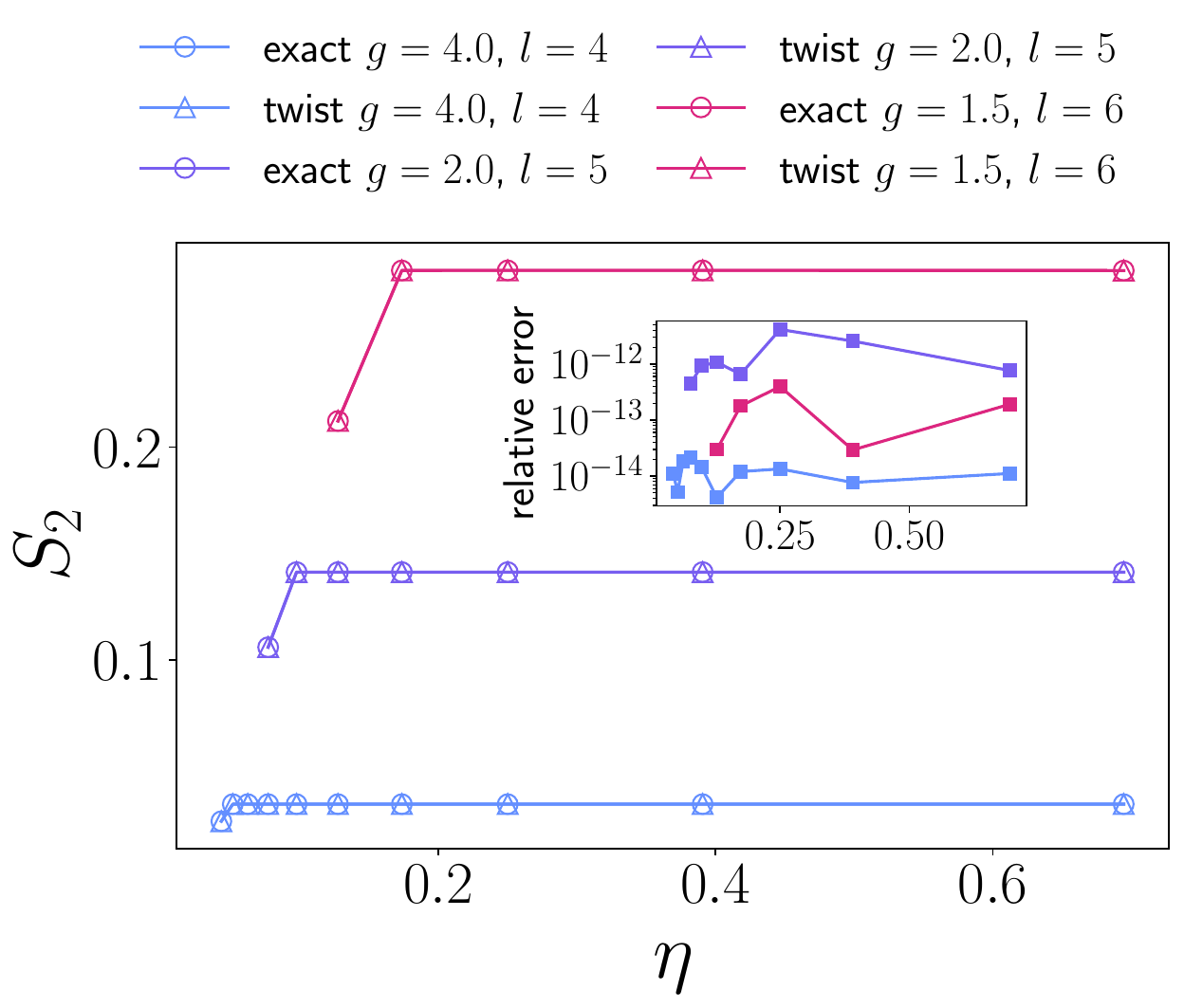}
    \caption{}
    \label{subfig:four_point_function}
    \end{subfigure}
    \caption{(a) Correlator of two twist operators as a function of their distance $r$. (b) Four-point function of twist field operators as a function of the ratio $\eta$. In both plots $L=120$ and the injectivity limit is assumed.}
    \label{fig:correlators_twist}
\end{figure*}

As highlighted in ref.~\cite{Coser:2013qda}, the R\'enyi entropy of $k$ 
disjoint 
intervals 
can be obtained from 
$2k$-point correlation functions of twist fields. Accordingly, the R\'enyi entropies associated with an arbitrary number of intervals can be obtained by means of multiple insertions of our twist operators, 
as shown in fig.~\ref{fig:correlators_twist}. In particular, in fig.~\ref{subfig:two_point_function} the two-point function of the twist operator is computed as a function of the distance between the two operator insertions $r$. This corresponds to the second R\'enyi entropy of an interval of length $r$ within the spin chain. The insertions are chosen symmetrically with respect to the center of the chain. As 
another 
example, fig.~\ref{subfig:four_point_function} presents 
the results of 
the calculation of a four-point function, corresponding to the entropy of two separate intervals, as a function of the ratio $\eta = (x_{12}x_{34}) / (x_{13}x_{24})$, where $x_{ij}=|x_i - x_j|$ and $x_i$ are the positions of the four operator insertions. In all cases, the results obtained from the twist operators in the injectivity limit are exact 
within 
numerical precision.

\begin{figure}[t!]
    \centering
    \includegraphics[width=\linewidth]{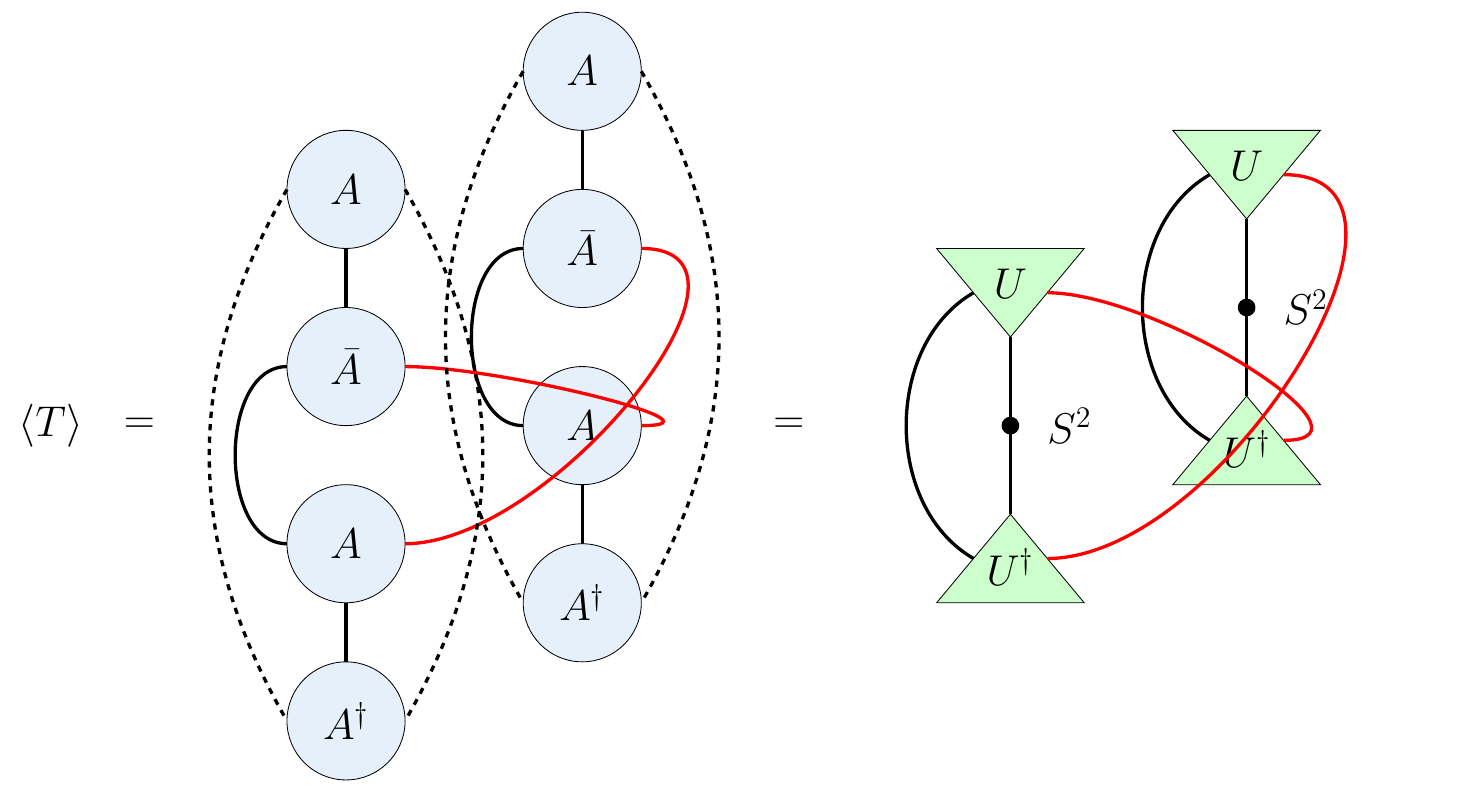}
    \caption{Expectation value of the twist operator in the center of orthogonality.}
    \label{fig:orthogonality_contraction}
\end{figure}

In the present section, we have worked exclusively in the injectivity limit. As a final remark, we note that there exists another setting in which the twist operator in eq.~\eqref{eq:twist_operator} exactly reproduces the R\'enyi entropy. Specifically, for a generic MPS, without assuming injectivity, eq.~\ref{eq:vev_twist_equal_Renyi} holds 
as long as
the tensor $A$ coincides with the center of orthogonality of the tensor network. In this case, the inversion of the MPS is not required. Indeed, when $A$ is chosen to be the orthogonality center, the contraction shown in fig.~\ref{subfig:virtual_contraction} is reproduced exactly by construction. This mechanism is illustrated in fig.~\ref{fig:orthogonality_contraction} for the case of a forward twist operator. By explicitly performing a singular-value decomposition, one can precisely match the action of the virtual swap induced by the twist operator, leading again to the equivalence between the R\'enyi entropy and the expectation value of the twist operator.

\section{Approximate schemes}
\label{sec:approximate_schemes}
Measuring entanglement entropies in many-body systems is intrinsically challenging for two main reasons. First, entropies are nonlinear functions of the quantum state as discussed before, and therefore cannot be obtained from the expectation value of a single observable, typically requiring multiple copies of the state or full reconstruction protocols. Second, they are \emph{non-local quantities}, as they depend on the reduced density matrix of an extended subsystem, making their direct experimental evaluation demanding. As a result, the required resources grow rapidly with system size.

In the previous sections, we analytically constructed local operators that, acting at the physical level, compute the R\'enyi entropy of a quantum spin chain. The construction is exact in the injectivity limit, or when the tensor considered is the center of orthogonality of the MPS. These observations naturally raise the question, whether such operators can be constructed also in experimental setups, to provide an efficient scheme to measure quantum entanglement. In what follows we address this question.

The setup we analyze in this section is the following. A twist operator is constructed from a reference quantum system of size $L$, and its expectation value is evaluated on a sequence of systems of increasing size. From a practical perspective, determining the twist operator on small chains is expected to be significantly less demanding than performing a full entanglement computation on large systems, especially in setups where the virtual Hilbert space is not directly accessible. This suggests a potentially advantageous strategy: one may construct an approximate twist operator from a relatively small system and then use it as a proxy to estimate the R\'enyi entropy in larger systems.

The central question is therefore the reliability of such an approximation as the system size increases. Since the construction is exact only under specific conditions (injectivity or orthogonality center), deviations are expected when these assumptions are relaxed. However, the quality of the approximation should depend sensitively on the physical properties of the state. In particular, for gapped systems with finite correlation length, local properties are expected to be weakly sensitive to boundary conditions and to the system size (at least for large enough $L$). In this regime, the twist operator constructed from a small chain should already capture the relevant local structure of the MPS tensors, leading to a stable and accurate estimate of the R\'enyi entropy, even when applied to larger volumes.

In the injectivity limit, and for a two-replica system, the twist operator we derived corresponds to a tensor acting on $2l$ physical indices
\begin{align*}
    T = \sum_{i_1,\dots,i_l,j_1,\dots,j_l} T_{i_1,\dots,i_l,j_1,\dots,j_l}\; \sigma^{i_1} \otimes \dots \otimes\sigma^{i_l}\otimes \\
    \sigma^{j_1}\otimes\dots\otimes\sigma^{j_l},
\end{align*}
where $i,j\in\{0,1,2,3\}$ 
and $\sigma^0 = I$.
The tensor $T_{i_1,\dots,i_l,j_1,\dots,j_l}$ is symmetric under any permutation of the $i$ indices, corresponding to the first replica, any permutation of the $j$ indices, corresponding to the second replica, and under simultaneous exchange of $i$ and $j$ indices. The number of independent components is therefore $\frac{1}{2}K(K+1)$ with $K = \binom{3+l}{3}$.

In the center-of-orthogonality construction, the twist operator acts non-trivially only on a single site per replica, leading to a substantial reduction in the number of parameters required to specify it. For $n=2$ replicas, the number of independent coefficients is $10$, a number that can be further reduced if additional information is available, for instance from symmetry constraints that enforce the expectation value of specific Pauli strings to be identically vanishing.

\begin{figure*}[t]
    \centering
    \begin{subfigure}{.32\linewidth}
    \includegraphics[width=1\linewidth]{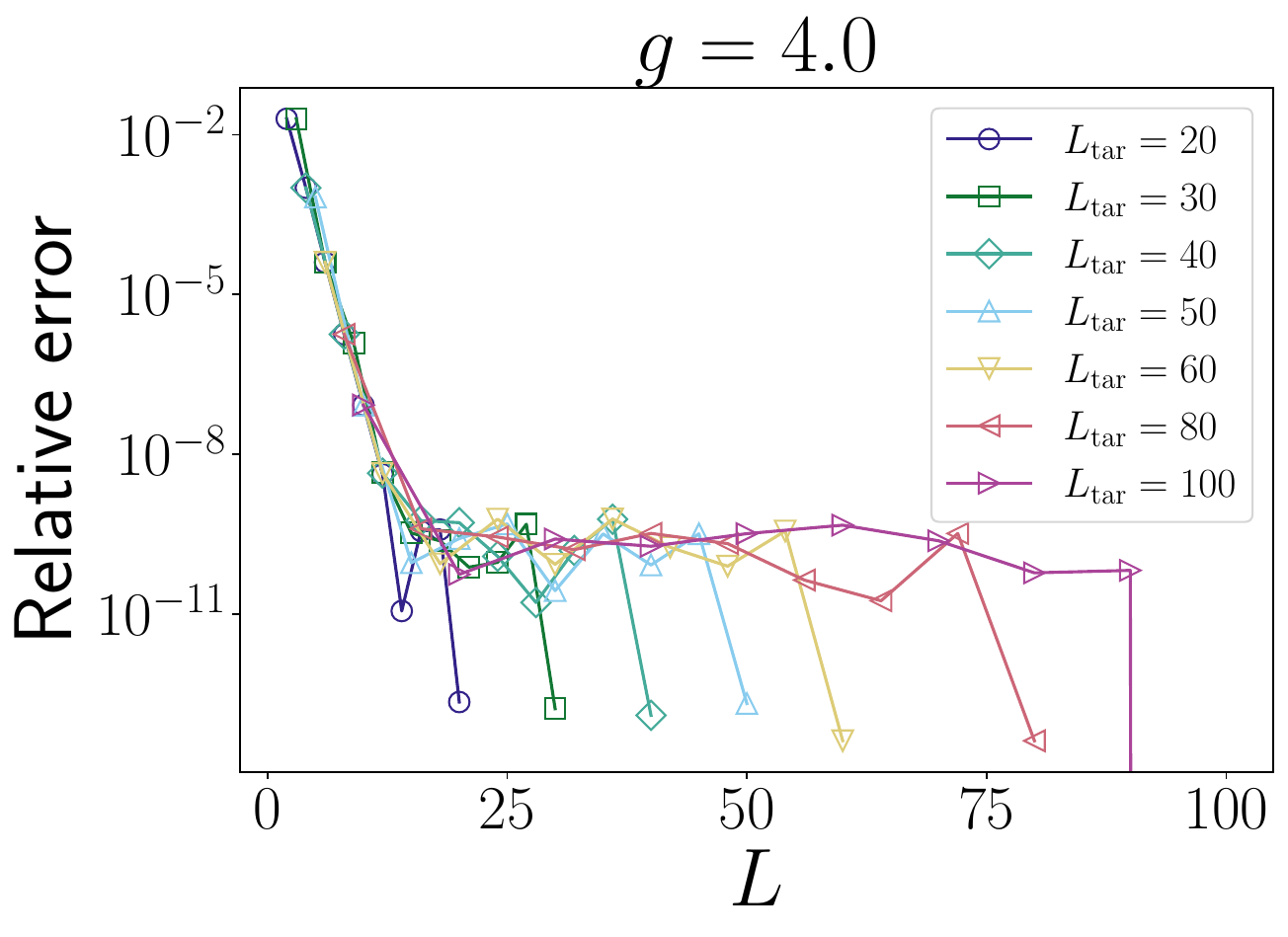}
    \caption{}
    \label{subfig:volume_transfer_4.0}
    \end{subfigure}
    \begin{subfigure}{.32\linewidth}
    \includegraphics[width=1\linewidth]{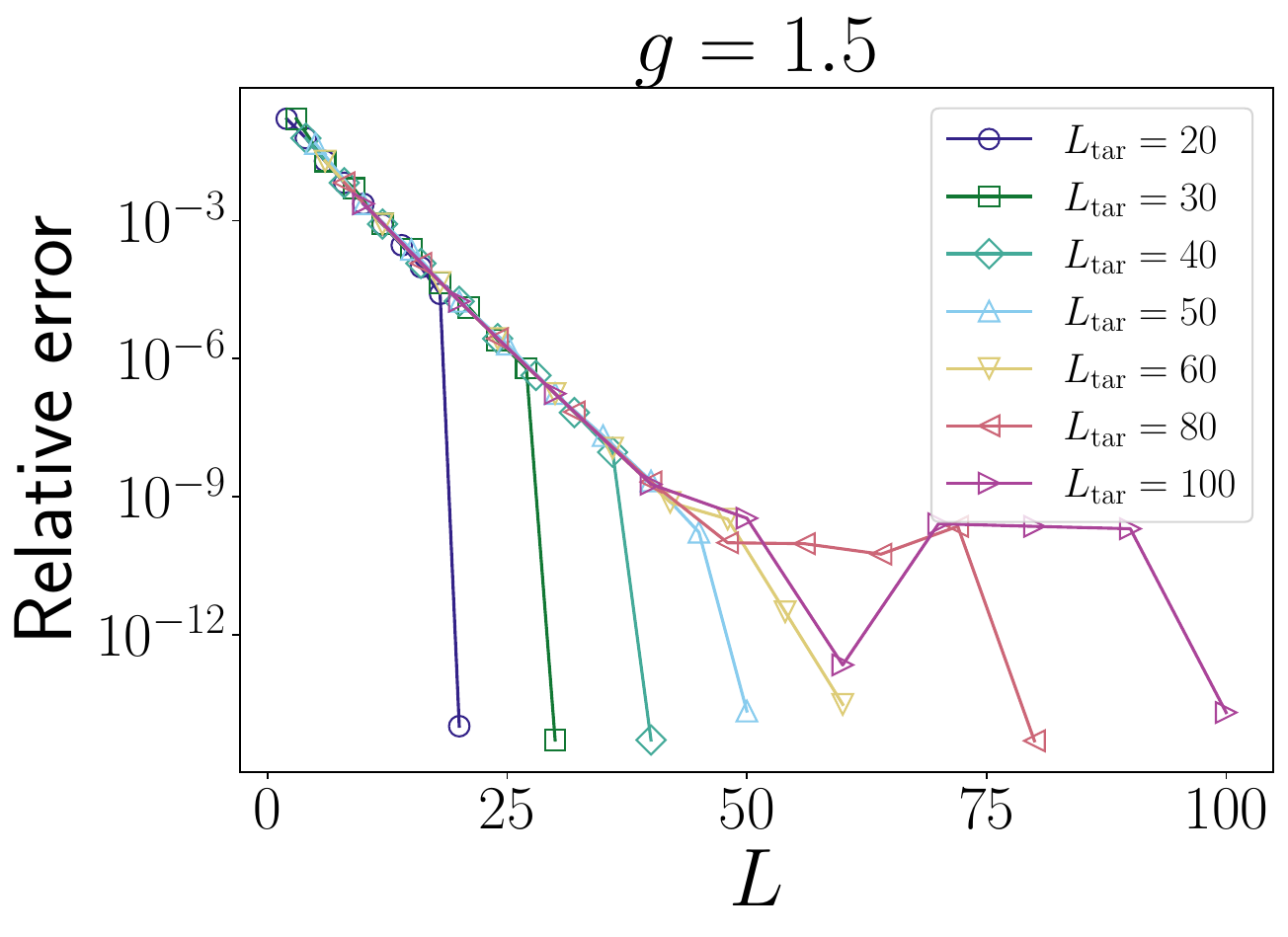}
    \caption{}
    \label{subfig:volume_transfer_1.5}
    \end{subfigure}
    \begin{subfigure}{.32\linewidth}
    \includegraphics[width=1\linewidth]{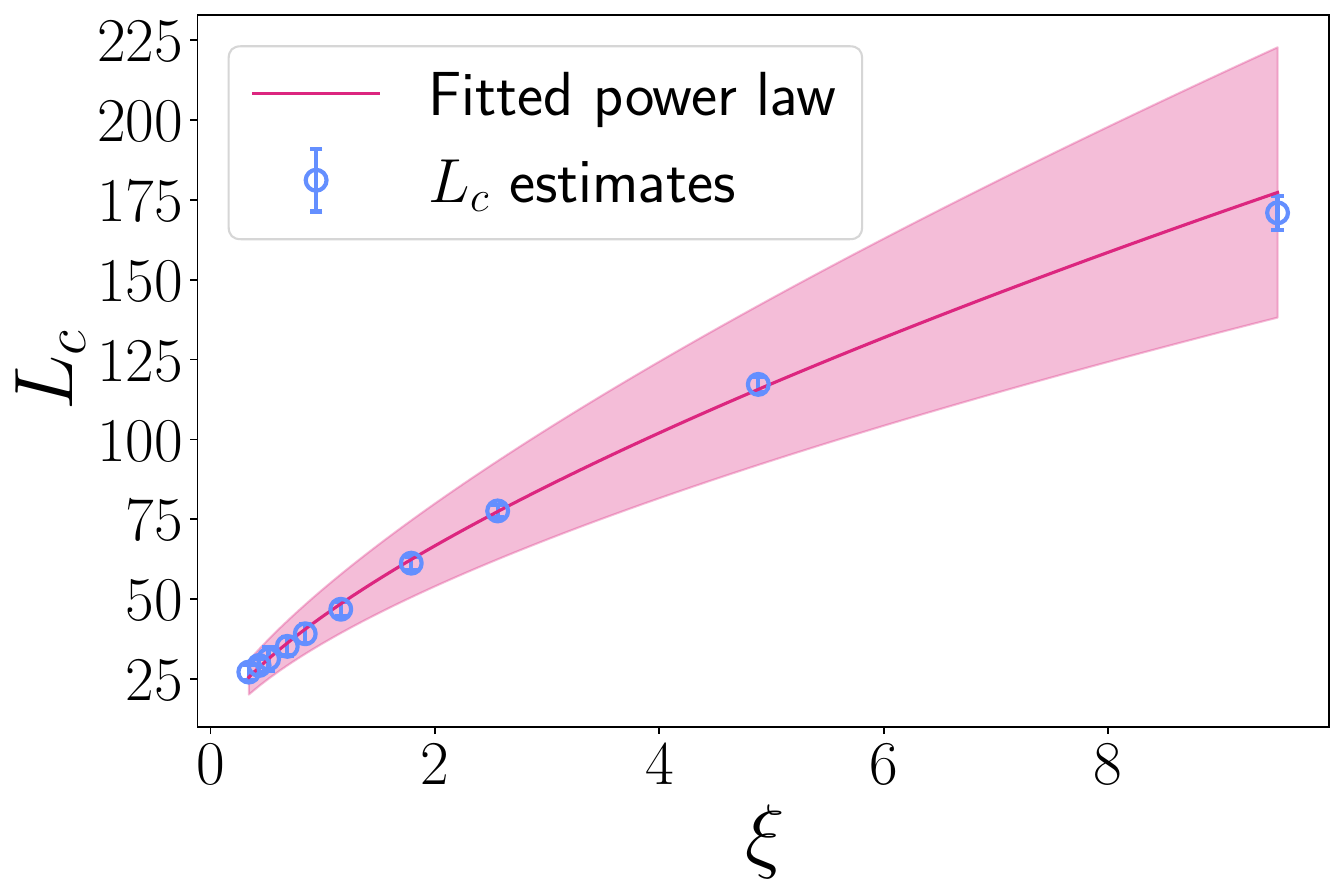}
    \caption{}
    \label{subfig:fit_Lc}
    \end{subfigure}
    \caption{(a) and (b): comparison between the exact half chain R\'enyi entropy computed in a chain of length $L_{\mathrm{tar}}$ and the expectation value of a twist operator $\langle T_{L} \rangle_{L_{\mathrm{tar}}}$. (c): values of $L_c$ as a function of the correlation length of the system, with the result of a fit to eq.~\eqref{eq:fit_Lc}.}
    \label{fig:volume_transfer}
\end{figure*}

In the following, we focus on the latter construction. Given a target chain length $L_{\mathrm{tar}}$, we consider a sequence of smaller systems with $L < L_{\mathrm{tar}}$, from which we extract the corresponding twist operators $T_L$. These are then evaluated on the target system, yielding expectation values $\langle T_L \rangle_{L_{\mathrm{tar}}}$.

In fig.~\ref{fig:volume_transfer}, we plot the relative error in the resulting estimate of the R\'enyi entropy with respect to the exact value. For fixed coupling $g$, the relative error decays exponentially with $L$, with a decay rate that is independent of $L_{\mathrm{tar}}$. For sufficiently large $L_{\mathrm{tar}}$, this enables a controlled extrapolation procedure: by computing $\langle T_L \rangle_{L_{\mathrm{tar}}}$ for increasing $L$, one can reliably estimate $S_2$ in the target system. However, this procedure is effective only when $L_{\mathrm{tar}}$ is larger than a critical threshold $L_c$. For $L_{\mathrm{tar}} < L_c$, the extrapolated value remains systematically biased. This threshold is expected to be set by the correlation length $\xi$, as for small $L$ the twist operators extracted from small systems fail to capture long-range correlations in the system: indeed, the twist field systematically underestimates the amount of entanglement in the system, i.e. $S_2^{\mathrm{exact}} \geq -\log\langle T_L  \rangle_{\Ltar}$.

This expectation is confirmed in fig.~\ref{subfig:fit_Lc}. For a range of values of $g$, we extract $L_c$ by fitting the exponential decay of the relative error and defining $L_c$ as the value of $L$ at which the relative error reaches a tolerance value, that we conventionally set to $10^{-10}$. The resulting values of $L_c$ are then fitted to the functional form
\begin{align}
L_c(\xi;A,\omega,k) = A \xi^{\omega} + k.
\label{eq:fit_Lc}
\end{align}
The correlation length $\xi$ is obtained in the thermodynamic limit via iDMRG simulations using TeNPy~\cite{Hauschild:2024bbz}. The fit is performed using a bootstrap procedure, yielding the estimates
\begin{align}
A = 37(4), \quad \omega = 0.68(5), \quad k = 8(4).
\end{align}
Note that $k$, which is expected to be nearly vanishing, is indeed compatible with zero within two standard deviations. The exponent $\omega$ may be related to the scaling dimension of the twist field. Further studies varying the number of replicas $n$ are required to clarify this connection.

\section{Experimental accessibility of the twist operator}
\label{sec:experimental_accessibility_of_the_twist_operator}

A key feature of the construction presented above is that the twist operator can be expressed as a local operator acting entirely on the physical Hilbert space, and can be decomposed into a linear combination of a small number of Pauli strings. 

A key feature of the construction presented above is that
the twist operator acts entirely on the physical Hilbert
space. Once expressed in a local operator basis, it can be
written as
\[
    T=\sum_{\alpha} a_{\alpha} P_{\alpha},
\]
where \(P_{\alpha}\) are Pauli strings acting on the two
replicas. Its expectation value is therefore obtained from
\[
    \langle T\rangle
    =
    \sum_{\alpha} a_{\alpha}\langle P_{\alpha}\rangle .
\]
Thus, after the coefficients \(a_{\alpha}\) have been
determined, the measurement of the Rényi entropy reduces
to the measurement of a finite set of local Pauli-string
expectation values.

Due to the symmetries of the construction, the number of independent coefficients required to specify the operator remains relatively small, even when the blocking length increases.

There are two natural ways of determining these
coefficients. The first is a classical calibration route:
one computes a tensor-network approximation to a small
reference system and constructs the corresponding twist
operator from the MPS tensors, as described above. The
resulting Pauli coefficients can then be used directly in
the experiment. The second possibility is an experimental
calibration route, in which the coefficients are inferred
from measurements on a small reference system by solving
the corresponding linear reconstruction problem. In both
cases, the key point is that the operator need only be
determined on a reference system whose size is controlled
by the correlation length.

The transferability demonstrated in fig.~\ref{fig:volume_transfer} provides the
basis for this protocol. In the transverse-field Ising
model, twist operators extracted from reference systems
of size \(L\) give increasingly accurate estimates of the
half-chain Rényi entropy of larger target systems. For
fixed coupling, the error decreases rapidly with the
reference size and becomes essentially independent of the
target size once \(L\) exceeds a characteristic scale \(L_c\).
The latter grows with the correlation length, as shown in
fig.~\ref{fig:volume_transfer} (c). This indicates that the physical twist operator
is controlled by the local structure of correlations around
the entanglement cut, rather than by the total volume of
the system.

From an experimental point of view, this provides a
reduction of the measurement problem. The conventional
replica protocol for \(S_2\) requires measuring a swap
operator acting on all sites in the region \(A\) \cite{Cardy:2011zz,abanin2012,daley2012,islam2015,kaufman2016}. By contrast,
the physical twist operator constructed here acts only on
a finite neighborhood of the boundary of \(A\), with a
size set by the the correlation length of the state rather than
by the volume of the subsystem.

The approach we propose is particularly well suited to platforms in
which local Pauli observables and few-body correlators
can be measured efficiently, such as Rydberg atom arrays \cite{barredo2016,browaeys2016,bernien2017a,scholl2021,semeghini2021,shaw2024},
trapped ions \cite{blatt2012,brydges2019,joshi2023,guo2024b,schuckert2025}, and superconducting qubits \cite{gong2021,kim2023,karamlou2024,andersen2025}. In practice,
one prepares two identical copies of the state and measures
the Pauli strings appearing in the decomposition of \(T\).
Grouping mutually commuting strings, or using randomized
measurement protocols, can further reduce the number of
measurement settings \cite{elben2020,brydges2019,huang2020}. The protocol does not require full
state tomography or reconstruction of the reduced density
matrix.

We emphasize that the twist operator is state-dependent:
it is determined by the tensor-network representation, or
by the calibrated local structure, of the reference state.
The advantage of the approach is therefore not that a
single universal observable measures the Rényi entropy
for all states, but rather that, for a given phase or family
of states with finite correlation length, a locally calibrated
operator can be transferred to larger systems and used to
estimate entanglement from a restricted set of local
measurements.

\section{Higher dimensions: PEPS}
\label{subsec:higher_dimensions_PEPS}
The key idea in 1D was to replace the physical swap used in the replica construction of R\'enyi entropies by an operator acting on the effective virtual degrees of freedom, and then approximate that action with a more local physical operator. In 2D one can still apply the same strategy, 
but the corresponding virtual twist is no longer point-like: it becomes an extended operator supported 
on the boundary of the region. The generalization of the above construction to higher-dimensional tensor networks, therefore, is not completely trivial.

In higher dimensions, the natural generalization of MPS is given by PEPS. Writing R\'enyi entropies still requires constructing replicated Riemann surfaces from copies of the double-layer contraction of the tensor network. R\'enyi entropies can again be expressed as expectation values of the swap operator acting on the physical degrees of freedom inside the region of interest, evaluated in the replicated system.

\begin{figure}
\includegraphics[width=.8\columnwidth]{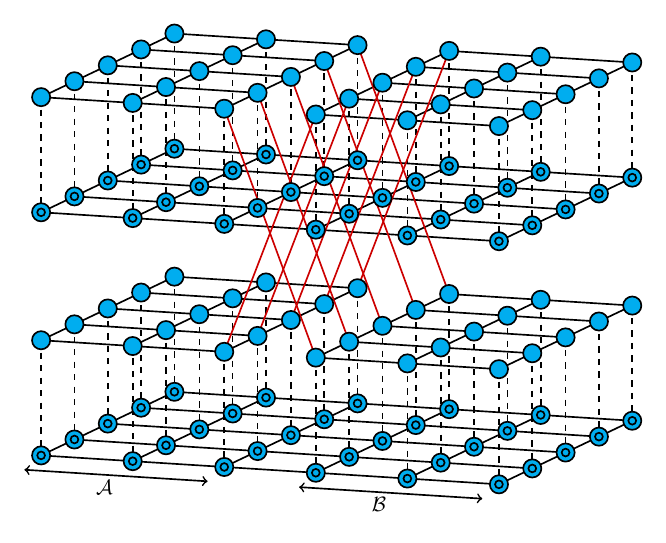}
\caption{R\'enyi  entropy of a region $\mathcal{A}$ in 2D PEPS, expressed as the expectation of the swap (solid red lines) of the virtual bond dimension along the boundary of $\mathcal{A}$. As in fig.~\ref{fig:virtual_real_contraction}, blue circles represent the ket, tensors with a circle inside the bra, solid lines are virtual bonds and dashed ones are physical bonds.}
\label{fig:virtual_twist}
\end{figure}

In the tensor-network representation of these replicated expectation values, one can still deform the contraction pattern analogously to what we 
discussed in the 1D case, 
and then identify a swap operator acting at the virtual level of the tensor network. However, an important difference is that, in 
2D, 
to compute the entanglement associated with a region $\mathcal{A}$,
the swap operator at the virtual level, and hence the corresponding twist fields, are necessarily extended objects: they are string-like operators acting on the boundary of the region $\mathcal{A}$, namely $\partial\mathcal{A}$, as shown for example in fig.~\ref{fig:virtual_twist}.

It is also important to notice that the two main properties we have been using for MPS, namely injectivity and the existence of a canonical form, are not as powerful in 2D. For example, different proposal have been pushed forward for the canonical form of a PEPS (see e.g. refs.~\cite{Acuaviva:2022lnc, Evenbly:2018njd} and the works cited therein). Also, contrary to standard PEPS whose exact contraction is exponentially hard (see e.g. ref.~\cite{Vasseur:2018gfy}), injective PEPS seems easier to deal with~\cite{Harley:2025gjs} and thus possibly less general. 

It is not straightforward, however, to carry out the same step as in the MPS case, namely to find a local operator acting on both replicas that mimics, even approximately, the action of the virtual swap. If such an object could be found, one would still obtain a substantial reduction in complexity, for example in the experimental determination of R\'enyi entropies. Indeed, while the original swap operator acts on the whole region $\mathcal{A}$, the twist-field operator would only act on $\partial \mathcal{A}$. This would provide a boundary-bulk advantage when evaluating its expectation value, compared with directly measuring the expectation value of the swap operator in the replicated system.

For example, in the simplest case of a PEPS for a spin-$1/2$ system, the size of the auxiliary Hilbert space of a block of $L\times L$ spins scales as\footnote{For notational convenience, we denote the PEPS bond dimension by $D$.} ${\cal H}_A = \mathbb{C}^{D^{4L}}$, while the size of the physical Hilbert space scales as ${\cal H}_P =  \mathbb{C}^{2^{L^2}}$. Putting numbers into these expressions, even for the smallest PEPS bond dimension, $D=2$, the smallest block for which one may expect injectivity is a $4\times 4$ patch of the lattice, involving $16$ spins, and leading to a coarse-grained PEPS with $D=8$. The resulting operator is therefore a $2^{16}\times 2^{16}$ operator, and even if one could invert it in the same spirit as in the MPS construction, its decomposition into operators acting on the block would become prohibitively large, involving of order $10^{18}$ terms. As a result, the strategy we used in one dimension is not the most suitable one in higher dimensions.

A more promising strategy is to consider partially contracted 2D tensor networks. For example, let us focus on the simplest case of an infinite 2D state described by an iPEPS \emph{Ansatz}, defined by the infinite repetition of an elementary tensor $T$, which we denote by $\ket{T}$. One possibility is to use a tensor-network renormalization approach \cite{Evenbly:2015ucs}, keeping track of the isometries used in the renormalization procedure, and then use them to obtain an effective elementary PEPS tensor $\tilde{T}$ with a small bond dimension $\tilde{D}$ describing a patch of the norm $\braket{T}$ that is large enough to be in the injective regime. One could then apply a strategy similar to the one used for MPS, inverting the $\tilde{D}^4$ map and obtaining a local operator that implements the virtual swap.

\begin{figure}[!hbt]
 \includegraphics[width=\columnwidth]{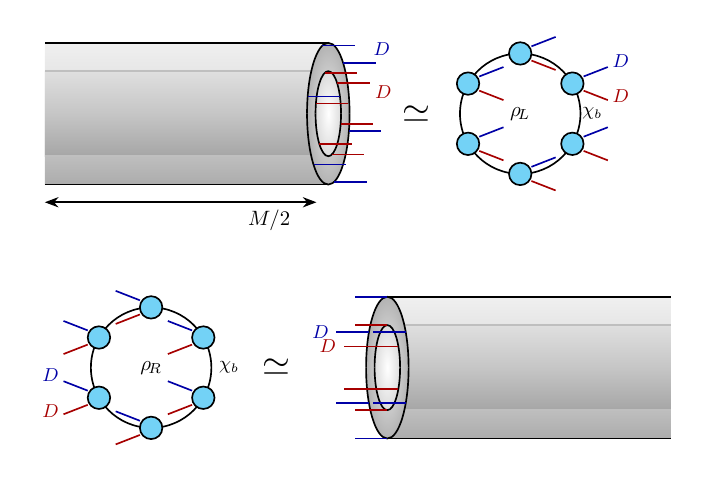}
 \caption{The ingredients for computing the reduced density matrix of the cylinder in terms of the boundary MPS $\rho_L$ and $\rho_R$.} 
 \label{fig:cyl_fixed_points}
 \end{figure}

An alternative construction is based on boundary contractions of $\braket{T}$. To make the discussion concrete, let us consider 2D systems defined on a cylinder, and assume that we want to compute the R\'enyi entropy of half of such a cylinder using a virtual swap, as shown in fig.~\ref{fig:virtual_twist}. We therefore contract a cylinder with compact direction of length $L$ and open direction of length $M\gg L$. We study the entropy of half of this cylinder, and assume for simplicity that $M/2\gg \xi$, where $\xi$ is the finite correlation length of the PEPS state. Importantly, we also assume that the contraction of $\braket{T}$ up to half of its length, $M/2$, from both sides, is well approximated by a boundary matrix-product operator (MPO) with finite bond dimension $\chi_b$, which we denote by $\rho_L$  for the contraction from the left of $M/2$, and by $\rho_R$ for the contraction from the right, both acting on the virtual Hilbert space $D^2L$ as shown in fig.~\ref{fig:cyl_fixed_points}. It is well known that the reduced density matrix of half of the cylinder has the same spectrum as $\rho=\sqrt{\rho_R}\rho_L\sqrt{\rho_R}$ \cite{Cirac:2011oss}.

The MPO tensors of $\rho=\sqrt{\rho_R}\rho_L\sqrt{\rho_R}$ have an auxiliary bond dimension $\chi_b$ that depends on the correlation length of the state. At this point, we are effectively back to an MPS problem, in which the R\'enyi entropy is given in terms of the expectation value of the tensor product of local swap tensors $S_D$ that exchange the bra and ket legs of $\rho$, each of dimension $D$, between two copies of the vectorized state $\ket{\rho}$ (note that there is no complex conjugation in the expectation value):
\begin{align}
\textrm{tr}\rho^2= \bra{\rho}\prod_{\otimes L} S_D\ket{\rho}.
\end{align}
Pictorially:
\begin{equation}\nonumber
\begin{tikzpicture}[baseline=(current bounding box.center),scale=0.4]
\def\dx{2.0}       
\def\Ntensors{6}   

\def\yTop{1.8}     
\def\yBot{-1.8}    

\def\legLen{0.9}
\def\legOff{0.30}

\def\arcHeight{1.1}

\def\rad{0.28}


\foreach \i in {0,...,4}{
  \pgfmathsetmacro{\xa}{\i*\dx}
  \pgfmathsetmacro{\xb}{(\i+1)*\dx}
  \draw[thick] (\xa, \yTop) -- (\xb, \yTop);
}

\pgfmathsetmacro{\xFirst}{0}
\pgfmathsetmacro{\xLast}{5*\dx}
\pgfmathsetmacro{\xMid}{2.5*\dx}
\pgfmathsetmacro{\yArcTop}{\yTop + \arcHeight}
\draw[thick] (\xFirst, \yTop) to[out=180-20, in=20] (\xLast, \yTop);

\foreach \i in {0,...,5}{
  \pgfmathsetmacro{\xc}{\i*\dx}
  \draw[thick] ({\xc - \legOff}, \yTop) -- ({\xc - \legOff}, {\yTop - \legLen});
  \draw[thick] ({\xc + \legOff}, \yTop) -- ({\xc + \legOff}, {\yTop - \legLen});
}


\foreach \i in {0,...,4}{
  \pgfmathsetmacro{\xa}{\i*\dx}
  \pgfmathsetmacro{\xb}{(\i+1)*\dx}
  \draw[thick] (\xa, \yBot) -- (\xb, \yBot);
}

\draw (1,1.3) node {\small $\chi_b$};
\draw (3.2,.4) node {\small \textcolor{red!80!black}{$D$}};
\draw (4.7,.4) node {\small \textcolor{red!80!black}{$D$}};
\draw (12.8,1.5) node {\small $\sqrt{\rho_R}\rho_L\sqrt{\rho_R}$};

\pgfmathsetmacro{\yArcBot}{\yBot - \arcHeight}
\draw[thick] (\xFirst, \yBot) to[out=-180+20, in=-20] (\xLast, \yBot);

\foreach \i in {0,...,5}{
  \pgfmathsetmacro{\xc}{\i*\dx}
  \draw[thick] ({\xc - \legOff}, \yBot) -- ({\xc - \legOff}, {\yBot + \legLen});
  \draw[thick] ({\xc + \legOff}, \yBot) -- ({\xc + \legOff}, {\yBot + \legLen});
}

\pgfmathsetmacro{\yTopLeg}{\yTop - \legLen}
\pgfmathsetmacro{\yBotLeg}{\yBot + \legLen}

\foreach \i in {0,...,5}{
  \pgfmathsetmacro{\xc}{\i*\dx}
  \draw[thick, color=red!80!black]
    ({\xc - \legOff}, \yTopLeg) -- ({\xc + \legOff}, \yBotLeg);
  \draw[thick, color=red!80!black]
    ({\xc + \legOff}, \yTopLeg) -- ({\xc - \legOff}, \yBotLeg);
}

\foreach \i in {0,...,5}{
  \pgfmathsetmacro{\xc}{\i*\dx}
  \fill[color=cyan] (\xc, \yTop) circle (\rad);
  \draw[thick]      (\xc, \yTop) circle (\rad);
  \fill[color=cyan] (\xc, \yBot) circle (\rad);
  \draw[thick]      (\xc, \yBot) circle (\rad);
}
\end{tikzpicture}
\end{equation}


The assumption $M/2\gg \xi$ implies that we do not expect the boundary MPS to change significantly upon contraction with one additional PEPS column.

\begin{figure*}[t]
\centering
\begin{subfigure}{.35\linewidth}
\includegraphics[width=.95\textwidth]{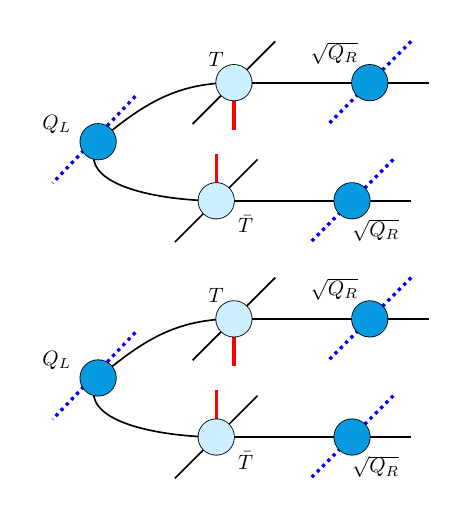}
\caption{}
\end{subfigure}
\hspace{1cm}
\begin{subfigure}{.35\linewidth}
\includegraphics[width=1.2\textwidth]{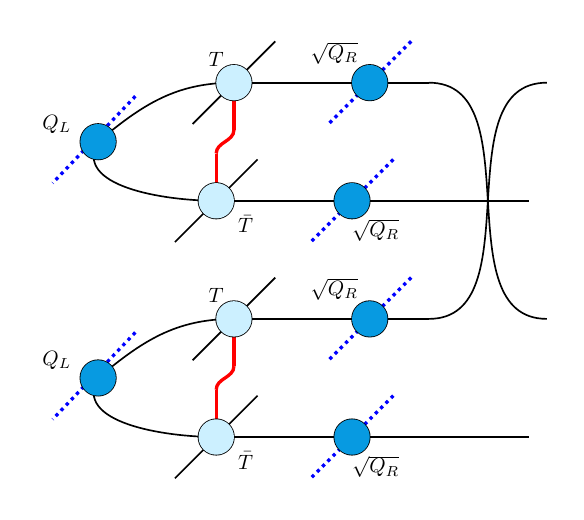}
\caption{}
\end{subfigure}
\caption{a) The  local part of $S_2$ we want to obtain by acting with the appropriate operator on the physical legs of the PEPS tensors in the two copies, black lines represent indices with dimension $D$, solid red line $d$ and dotted blue $\chi$  defines $\ket{L_T}$, a vector in the tensor product Hilbert space of all its open legs. 
b) Upon acting with $I\otimes S_T$ on all the physical legs and the virtual legs of the two copies, we obtain the local tensor network which provides $S_2$.
\label{fig:local_piece_S2}}
\end{figure*}

For simplicity we consider here a PEPS with bond dimension $D=2$, in such a way that adding a single column of uncontracted PEPS tensors could be sufficient to find the local representation of the twist field. We will also work locally, and detach a piece of the TN encoding for example $S_2$, as show in fig. \ref{fig:local_piece_S2}, where $Q_L$ and $Q_R$ are the MPO tensors of the $\rho_L$ and $\rho_R$.

\begin{figure*}[t!]
\includegraphics[width=\textwidth]{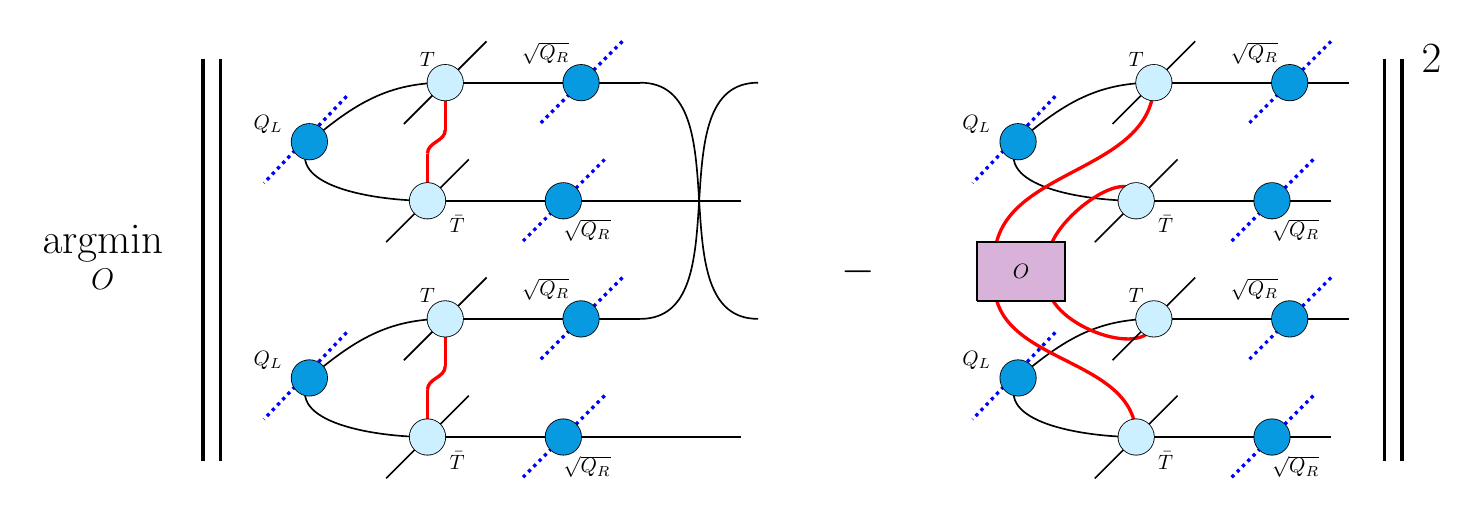}
\caption{The local operator encoding the twist field \(O_T\) is obtained by minimizing the distance between the vector produced by the physical operator, \(O_T|L_T\rangle\), and the vector produced by the virtual swap, \((I\otimes S_D)|L_T\rangle\), as described in eq.~\eqref{eq:argmin_optimization}.
\label{fig:def_O_T}}
\end{figure*}

Now we would like to identify operators acting on the physical legs of the two copies, the red ones in fig.~\ref{fig:local_piece_S2}, whose action reproduces, as accurately as possible, the action of the swap on the virtual legs, the black ones. This can be formulated as a variational problem in the space of local physical operators.

Let \(|L_T\rangle\) denote the vector obtained from the local tensor network before the virtual swap is applied, and let
\[
    |L_T^{\rm sw}\rangle = (I\otimes S_D)|L_T\rangle
\]
be the corresponding vector after applying the swap \(S_D\) on the virtual legs. We then look for a physical operator \(O_T\), acting only on the physical legs of the two replicas, such that
\[
    O_T |L_T\rangle \simeq |L_T^{\rm sw}\rangle .
\]
Equivalently, \(O_T\) can be defined as the solution of the least-squares problem
\begin{align}
    O_T = \underset{O}{\mathrm{argmin}}\,
    \left\| O|L_T\rangle - (I\otimes S_D)|L_T\rangle \right\|^2 ,
\label{eq:argmin_optimization}
\end{align}
possibly with the additional constraint \(O=O^\dagger\), if one wants the resulting twist operator to be directly measurable as an observable.

In practice, we restrict \(O_T\) to a finite operator basis on the physical Hilbert space. For spin-\(1/2\) systems, a natural choice is the basis of Pauli strings,
\[
    O_T = \sum_\alpha a_\alpha P_\alpha ,
\]
where \(P_\alpha\) denotes a tensor product of Pauli matrices acting on the physical legs of the two replicas. Substituting this expansion into eq.~(6) gives a finite-dimensional linear least-squares problem for the coefficients \(a_\alpha\). Defining
\[
    |v_\alpha\rangle = P_\alpha |L_T\rangle,
    \qquad
    |w\rangle = (I\otimes S_D)|L_T\rangle ,
\]
the normal equations read
\[
    \sum_\beta G_{\alpha\beta} a_\beta = b_\alpha ,
    \tag{7}
\]
with
\[
    G_{\alpha\beta} = \langle v_\alpha | v_\beta\rangle,
    \qquad
    b_\alpha = \langle v_\alpha | w\rangle .
    \tag{8}
\]
If the Gram matrix \(G\) is singular or ill-conditioned, the solution can be obtained using a Moore--Penrose pseudoinverse, or regularized by adding a small Tikhonov term,
\[
    a = (G+\lambda I)^{-1}b .
    \tag{9}
\]
This procedure gives the best approximation to the virtual twist within the chosen local operator subspace.

While in this section we developed the theoretical setup to access the twist field from 2D tensor network states, explicit simulations are expected to be more costly than in 1D, and numerical results will be presented elsewhere.

\section{Conclusions}
\label{sec:conclusions}

In this work, we have constructed local physical operators that reproduce the action of twist fields in tensor network states, providing direct access to R\'enyi entropies through expectation values of operators acting entirely on the physical Hilbert space. Our construction establishes an explicit correspondence between virtual swap operations, which are naturally defined at the level of auxiliary tensor network degrees of freedom, and experimentally accessible operators expressed in terms of physical observables.

We have shown that, in the injectivity limit or when the tensor is chosen at the center of orthogonality, the expectation value of the twist operator exactly reproduces the R\'enyi entropy. This provides a concrete and operational framework to compute entanglement measures within tensor network descriptions, without requiring direct access to the virtual degrees of freedom. Our numerical tests in the transverse-field Ising model confirm the validity of the construction and exhibit rapid convergence to the exact result as the injectivity scale is approached.

Furthermore, we demonstrated that twist operators constructed from relatively small reference systems can be reliably applied to larger systems. As shown in fig.~\ref{fig:volume_transfer}, once the reference system exceeds a characteristic size set by the correlation length, the twist operator becomes effectively independent of the total system size and provides accurate estimates of entanglement properties in significantly larger systems. This transferability reflects the local nature of the tensor network description and highlights the robustness of the construction.

An important consequence of our results is the experimental accessibility of the twist operator. Since it admits a decomposition in terms of a finite number of local Pauli operators, its expectation value can be reconstructed from a limited set of local measurements. This enables a direct estimate of R\'enyi entropies in experimental quantum simulators such as Rydberg atom arrays, trapped ions, and superconducting qubit platforms, without requiring full state tomography or reconstruction of reduced density matrices.

Our results are not limited to 1D systems: by means of PEPS, one can variationally construct operators acting on the border of the subsystem $\mathcal{A}$ whose expectation value approximate the R\'enyi entropy. Further analysis on the scalability of such operators with the system size, as well as the scaling in the number of Pauli strings involved, will be the subject of future works.

More broadly, our work provides a scalable and experimentally feasible framework to probe entanglement in quantum many-body systems using tensor network methods. Possible extensions include a generalization to higher-dimensional tensor network states, as well as applications to quantum simulation experiments. These results open new avenues for bridging tensor network theory and experimental quantum many-body physics through the direct measurement of entanglement observables.

\section*{Acknowledgments} A.B. thanks Anthony Gandon and Alessio Negro for useful discussions. A.B. was supported by the Deutsche Forschungsgemeinschaft (DFG, German Research Foundation) as part of the CRC 1639 NuMeriQS -- project no.\ 511713970 and under Germany’s Excellence Strategy -- Cluster of Excellence ``Color meets Flavor'' (CmF) EXC 3107 -- 533766364.
P.S. acknowledges funding from the Spanish Ministry for Digital Transformation and the Civil Service of the Spanish Government through the QUANTUM ENIA project call - Quantum Spain, EU, through the Recovery, Transformation and Resilience Plan – NextGenerationEU, within the framework of Digital Spain 2026.  L.T.  acknowledges the support from the Proyecto Sin\'ergico CAM Programa TEC-2024/COM-84 QUITEMAD-CM, the CSIC Research Platform on Quantum Technologies PTI-001, from the Grant TED2021-130552B-C22 funded by MCIN/AEI/10.13039/501100011033 and by the ``European Union NextGenerationEU/PRTR'', from the grant PID2024-160172NB-I00 funded by MICIU/AEI/10.13039/501100011033 and by FEDER, UE.

\bibliographystyle{quantum}
\bibliography{references}

\end{document}